\documentclass[conference]{IEEEtran}
\IEEEoverridecommandlockouts

\usepackage{cite}
\usepackage{amsmath,amssymb,amsfonts}
\usepackage{algorithmic}
\usepackage{graphicx}
\usepackage{textcomp}
\usepackage{xcolor}
\def\BibTeX{{\rm B\kern-.05em{\sc i\kern-.025em b}\kern-.08em
    T\kern-.1667em\lower.7ex\hbox{E}\kern-.125emX}}

\usepackage{lipsum}

\begin{document}

\title{Analyzing 16,193 LLM Papers for Fun and Profits}

\author{\IEEEauthorblockN{Zhiqiu Xia$^{*}$, Lang Zhu$^{*}$, Bingzhe Li$^{\dagger}$, Feng Chen$^{\dagger}$, Qiannan Li$^{\ddagger}$, Chunhua Liao$^{\S}$, Feiyi Wang$^{\P}$, and Hang Liu$^{*}$}
\IEEEauthorblockA{
\{zx283, lz529, hl1097\}@scarletmail.rutgers.edu, \\ \{bingzhe.li, feng.chen\}@utdallas.edu, qnli@ucdavis.edu, liao6@llnl.gov, fwang2@ornl.gov \\
$^{*}$Rutgers, The State University of New Jersey, New Brunswick, NJ, USA \\
$^{\dagger}$The University of Texas at Dallas, Richardson, Texas, USA \\
$^{\ddagger}$University of California, Davis, Davis, CA, USA \\
$^{\S}$Lawrence Livermore National Laboratory, Livermore, CA, USA \\
$^{\P}$Oak Ridge National Laboratory, Oak Ridge, TN, USA
}
}

\maketitle

\begin{abstract}
Large Language Models (LLMs) are reshaping the landscape of computer science research, driving significant shifts in research priorities across diverse conferences and fields. This study provides a comprehensive analysis of the publication trend of LLM-related papers in 77 top-tier computer science conferences over the past six years (2019–2024). 
We approach this analysis from four distinct perspectives: (1) We investigate how LLM research is driving topic shifts within major conferences. (2) We adopt a topic modeling approach to identify various areas of LLM-related topic growth and reveal the topics of concern at different conferences. (3) We explore distinct contribution patterns of academic and industrial institutions. (4) We study the influence of national origins on LLM development trajectories.
Synthesizing the findings from these diverse analytical angles, we derive ten key insights that illuminate the dynamics and evolution of the LLM research ecosystem.
\end{abstract}


\section{Introduction}

The recent rapid evolution of LLMs has significantly changed the research trend, especially in the domain of natural language processing (NLP). Over the past few years, LLMs such as BERT~\cite{devlin_bert_pre_training}, GPT-3~\cite{brown_language_model_are}, Llama~\cite{touvron_llama_open_and} and their derivatives have demonstrated unprecedented capabilities in generating, understanding, and adapting to human language. These advancements have impacted not only core NLP domains but also extended their influence across a wide array of interdisciplinary applications, including computer vision, robotics, systems, and so on. LLMs have shown a paradigm shift, altering the research trajectories of various fields and conferences by applying LLMs. The growing availability of LLMs with increasing computational resources, datasets, and evaluations has spurred an exponential increase trend in the research community. This trend can be reflected in the increasing number of LLM-related publications from all kinds of leading computer science conferences.

\subsection{Related Work \& Motivations}

Consequently, bibliometrics—the quantitative examination of publication patterns, research trends, and topical emphases—offers valuable insights into the evolution of scientific disciplines. Numerous studies have analyzed publication trends, collaboration networks, and thematic shifts across various research domains. For instance, \cite{Dong-KDD17-century} analyzed 89 million scholarly publications from 1900 to 2015, revealing a notable transition towards collaborative research efforts over time. Within the NLP community specifically, recent studies have examined citation patterns and paradigm shifts in research topics. \cite{mohammad-ACL2020-examining} analyzed citation trends within NLP literature, while \cite{pramanick-EMNLP2023-diachronic} developed a causal discovery and inference-based framework to systematically trace the evolution and detect paradigm shifts in NLP research.

More recently, the rise of LLMs has become a focal point in bibliometric analyses. For example, \cite{Fan-TIST2024-review} conducted a comprehensive study covering over 5,000 publications related to LLM research from 2017 to early 2023, providing detailed insights into the evolution of foundational models, interdisciplinary application trends, and collaborative patterns. Similarly, \cite{movva_topics_authors_and} analyzed 16,979 arXiv publications on LLMs, highlighting shifts in topics, authorship patterns, and institutional collaborations. Their findings pointed to an increasing disciplinary shift toward societal implications of LLMs and a growing influx of researchers from outside traditional NLP fields.

However, the scope of these previous bibliometric studies is inherently limited by their reliance on specific datasets, such as the ACL Anthology corpus or the arXiv repository. While these datasets provide valuable and specialized coverage, they tend to limit the bibliometric analysis to a particular field or preprint platforms, potentially missing broader disciplinary insights and introducing biases toward certain topics or trends. 

\subsection{Contributions}
This study dismantles these limitations by providing a comprehensive analysis of LLM-related research trends within virtually all major top-tier computer science (CS) conferences. By examining more than 160k+ published papers in 77 conferences across diverse subfields over the past six years, we identify 16,193 LLM-related papers capturing ten valuable insights. This paper's contributions are as follows:

\begin{enumerate}
    \item \textbf{LLM-related papers' publication trend of top conferences: }We examine the LLM research landscape across leading CS conferences. This includes conferences spanning artificial intelligence (AI), systems, theory, and interdisciplinary areas that previous works have not comprehensively addressed.
    \item \textbf{Topics shifting driven by LLMs: }We study topics shifting by topic modeling to provide comprehensive observations of topics in these subfields. This result echoes the recent societal concerns and hot topics in the research community. 
    \item \textbf{Impact of academic and industrial contributions on LLM research: } We explore how the academic and industry institutions shape the LLM research in these conferences. By analyzing publication trends, we study which institutions have the highest output and impact in the field of LLMs. 
    \item \textbf{The role of nations in shaping research trends: }We analyze how nations contribute to research development in different fields. This includes examining national influences on topic distribution.
\end{enumerate}

The remainder of this paper is organized as follows: Section~\ref{sec:dataset_and_methods} details the datasets and analytical methods employed in this study. Subsequently, we present and analyze our results in depth in Section~\ref{sec:resutls_and_analysis}, building upon these findings to culminate in ten key insights that characterize the evolving landscape of LLM research. Finally, Section~\ref{sec:conclusion} concludes the paper by summarizing our work and its implications.

\section{Dataset and Methods} \label{sec:dataset_and_methods}

\subsection{Identifying LLM-related Papers}
Our dataset contains {168,331} papers from 2019 to 2024 of 77 prestigious computer science and engineering conferences according to CSRankings~\cite{CSRankings}. These 77 conferences encompass fields from algorithms to emerging areas of computer science and engineering, covering four main avenues: AI, systems, theory, and interdisciplinary areas.

\begin{figure}[htbp]
\centering
\begin{minipage}[t]{\linewidth}
  \centering
  \includegraphics[width=.95\textwidth]{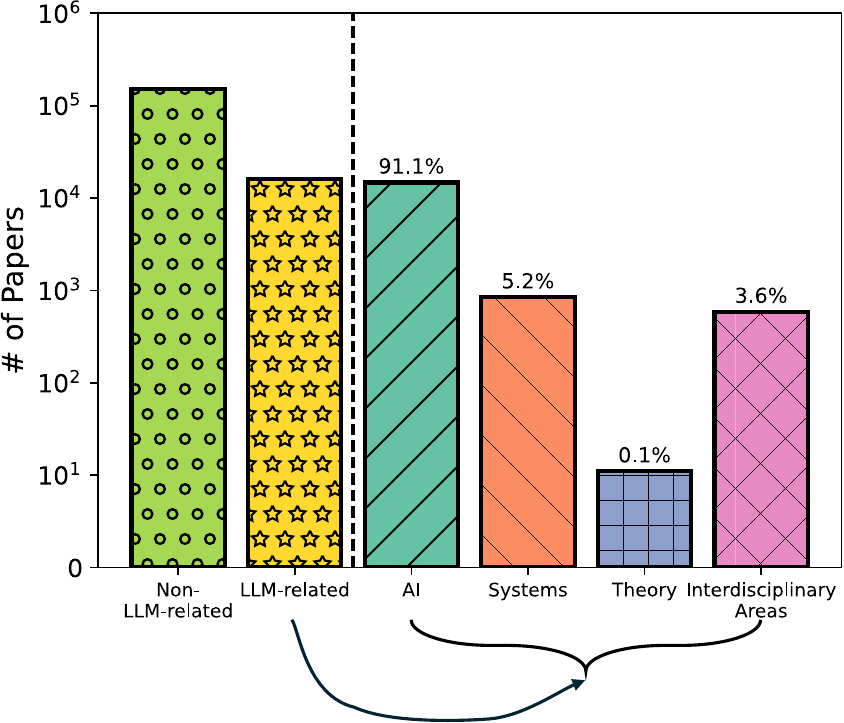}
  \vfill
    \caption{Number of LLM-related papers in each area.}
    \label{fig:dataset_stats}
\end{minipage}
\end{figure}

To identify LLM-related papers, \cite{movva_topics_authors_and} composes a set of LLM-related keywords and runs keyword matching throughout the title and abstract across all the papers in the dataset to pick LLM-related papers from them. However, we find an LLM-related paper might not have those keywords in its title or abstract. Further, it is challenging to pre-define a comprehensive set of keywords for LLM, which is a rapidly evolving field. Instead, we leverage the capabilities of LLMs themselves to identify LLM-related papers. Specifically, we define the criteria for what constitutes an LLM-related paper in the prompt. Subsequently, we deploy a local \textit{Llama3.1-8B-Instruct}~\cite{grattafiori_the_llama_3} model to determine whether a given paper is related to LLMs or not based on its title and abstract.

Our selection process yields 16,193 LLM-related papers, with the numbers of papers in each area shown in Fig.~\ref{fig:dataset_stats}. 
Table~\ref{tab:dataset_stats} further shows the \# of papers published in each year during 2019 - 2024.
In 2019, there are only 503 LLM-related papers, indicating that research in this area was still in its early stages. And the number doubles in 2020 and climbs further in 2021. Beginning in 2022, there was a sharp increase in the number of LLM papers. The most significant surge occurred in 2024, with 7,109 LLM-related papers, representing an increase of 3,255 
papers compared to 2023. This dramatic growth underscores how rapidly LLM is taking a central role in advancing CS research.

\begin{table}[tbp]
\centering
\caption{Number of LLM-related papers from 2019 to 2024}
\label{tab:dataset_stats}
\resizebox{\columnwidth}{!}{%
\begin{tabular}{l|l|l|l|l|l|l}
\hline
Year   & 2019 & 2020 & 2021 & 2022 & 2023 & 2024 \\ \hline
Number & 503  & 1105  & 1595 & 2027 & 3854 & 7109 \\ \hline
\end{tabular}%
}
\end{table}

\subsection{Topic Modeling} \label{sec: topic_modeling}
To better understand the research directions and focus areas of LLM-related papers, we apply \textit{topic modeling} on the entire corpus of LLM-related papers. Topic modeling~\cite{ABDELRAZEK2023102131} is an unsupervised machine-learning technique that identifies the latent thematic structure within a collection of documents. We use this tool to extract and analyze the various research directions and focus areas present in LLM-related research.

While traditional topic modeling methods, such as Latent Dirichlet Allocation (LDA)~\cite{blei_latent_dirichlet_allocation} and Non-Negative Matrix Factorization (NMF)~\cite{lee_algorithms_for_nonnegative}, are widely used for this purpose, we leverage an advanced LLM-based approach for topic modeling (in part, inspired by~\cite{movva_topics_authors_and}). Specifically, we first utilize \textit{INSTRUCTOR-XL}~\cite{su_one_embedder_any} to get semantic text embeddings of the abstract for each paper, with the prompt ``Represent the Computer science abstract for clustering''.
These embeddings are then clustered using Ward's agglomeration clustering\cite{murtagh_wards_hierarchical_agglomerative}, producing 50 clusters. Finally, to identify a representative topic for each cluster, we queried \textit{Gemini-2.0-Flash}~\cite{geminiteam_gemini}, which provides a concise and meaningful label for each cluster.

\subsection{Identity Affiliation and Nation}
To determine the affiliation of each paper, we adopt a multi-step approach. First, we utilize the \textit{OpenAlex} API~\cite{priem_openalex_a_fullyopen}, which provides affiliation information for a certain portion of the papers. For papers where affiliation information is unavailable through OpenAlex, we extract the affiliation details from the first page of the paper’s full text using \textit{Llama3.1-8B-Instruct}. If \textit{Llama3.1-8B-Instruct}, unfortunately, is unable to identify the affiliation accurately, we manually verify and extract the information through additional searches.

Once all affiliations are collected, we standardize and consolidate variations in affiliation names (e.g., ``Massachusetts Institute of Technology'' and ``MIT'') using \textit{Gemini-2.0-Flash}, leveraging its ability to handle long-context inputs effectively. Finally, we query \textit{Llama3.1-8B-Instruct} to determine the nation associated with each affiliation, ensuring accurate mapping between institutions and their corresponding countries.

\section{Results and Insights}~\label{sec:resutls_and_analysis}
\vspace{-.1in}
\subsection{Conference Analysis}

\begin{figure}[t]
\centering
\includegraphics[width=\columnwidth]{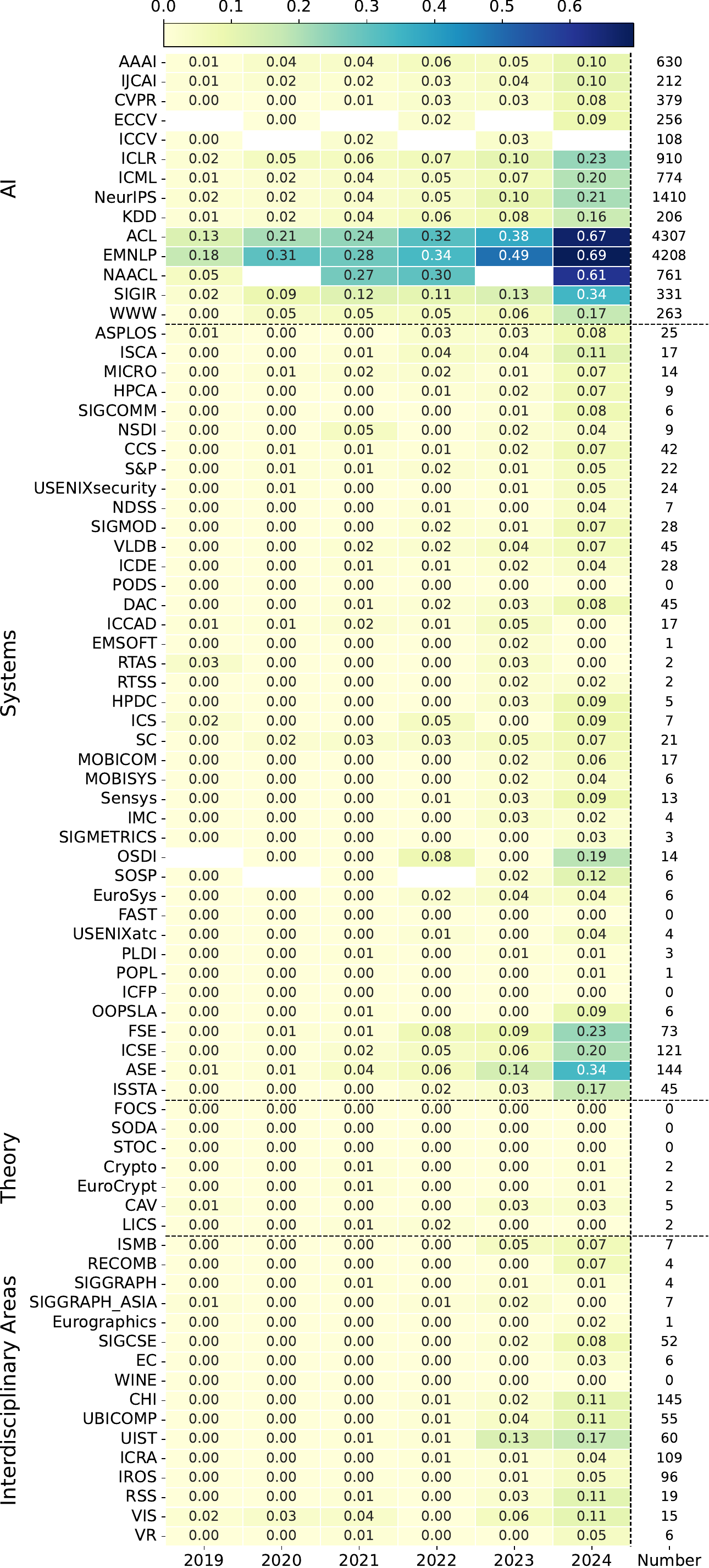}
\caption{LLM-related paper ratio across different conferences.}
\label{fig:figure_llm_related}
\end{figure}

Fig.~\ref{fig:figure_llm_related} shows the dynamics in \# of papers in per conference year dynamics for all the 77 conferences. Among the LLM-related papers, most are concentrated in the field of AI, whose \# of papers is an order of magnitude higher. A substantial portion of the study also focused on systems research and interdisciplinary applications. In contrast, very few LLM-related papers appear in the theoretical territory. Such a result is unsurprising, as LLM research originates from the NLP domain within AI.

\textbf{Insight \#1. In the AI field, LLM-related research exhibits a dominance in NLP ($>$ 60\%), substantial presence in ML ($>$ 20\%), and rising tractions in computer vision ($\sim$10\%).} Conferences focused on NLP, such as ACL, EMNLP, and NAACL, show the highest proportion of LLM-related papers, with values exceeding 60\% by 2024. Machine learning conferences, including NeurIPS, ICLR, and ICML, and information retrieval conferences, including SIGIR and WWW, also observe a steady increase, reaching around 20\% LLM-related papers by 2024. Besides, computer vision conferences like CVPR and ECCV in 2024 show an increase in the proportion of LLM-related papers, highlighting the growing popularity of multi-modal LLMs.

\textbf{Insight \#2. In the Systems area, architecture for LLM (ISCA), system for LLM (OSDI), and LLM for software engineering (ASE) propelled the growth in LLM papers.} Conferences generally show ratios rising to around 10\% by 2024, reflecting the growing interest in designing more efficient and scalable systems to support LLMs, as seen in venues like OSDI/SOSP. Additionally, software engineering conferences such as FSE, ICSE, ASE, and ISSTA, which focus on topics like code generation, exhibit notable LLM-related research activity, further highlighting the expanding influence of LLMs in the software engineering domain.

The Theory area has the lowest representation of LLM-related research across all categories, indicating that theoretical exploration of LLMs remains to be unveiled.

\begin{figure*}[tbp]
    \centering
    \begin{minipage}[t]{.32\textwidth}
        \centering
        \includegraphics[width=\textwidth]{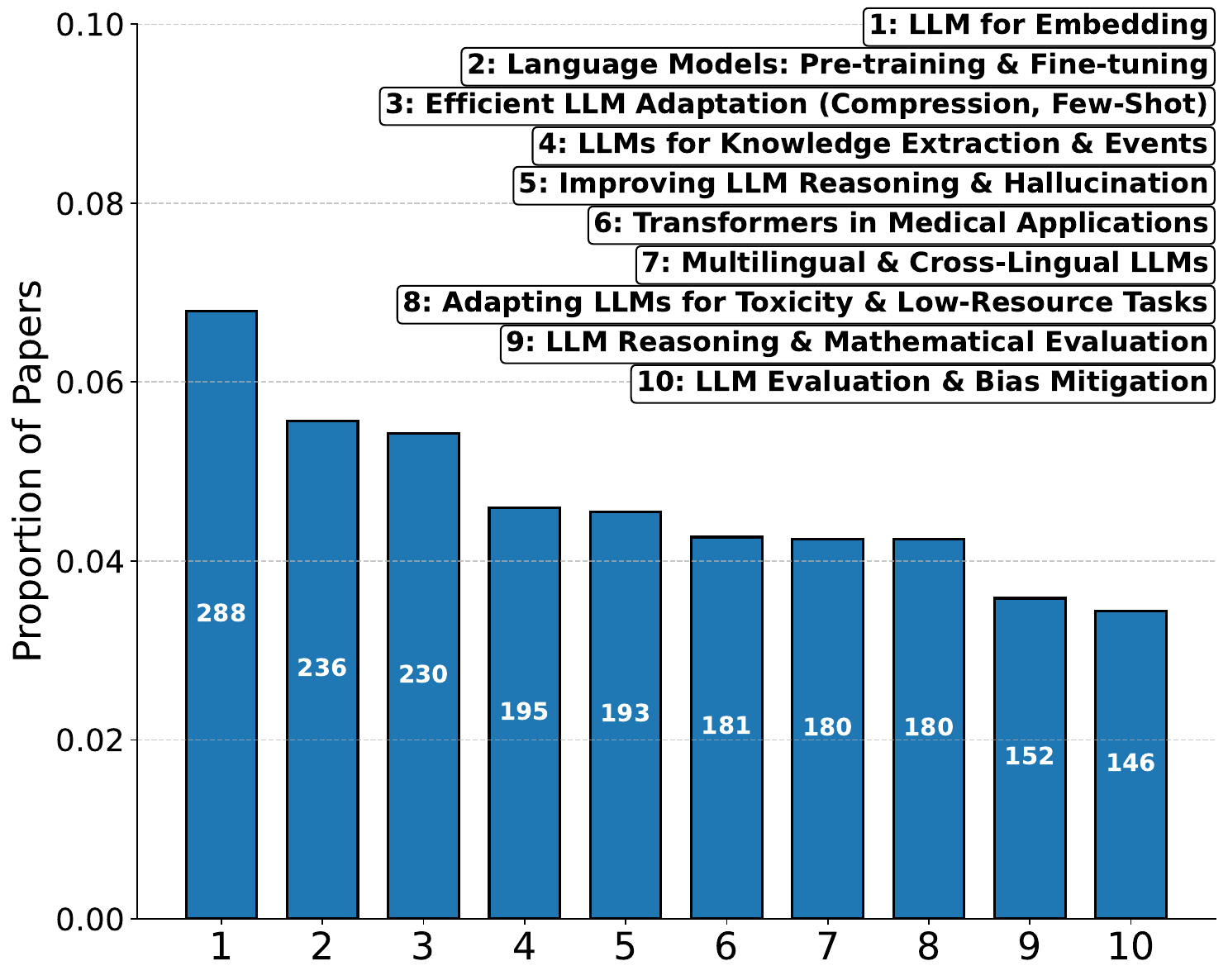}
        \vspace{-0.3in}
        \caption{Top 10 topics in ACL.}
        \label{fig:ACL}
    \end{minipage}
    \hspace{0.03in}
    \begin{minipage}[t]{.32\textwidth}
        \centering
        \includegraphics[width=\textwidth]{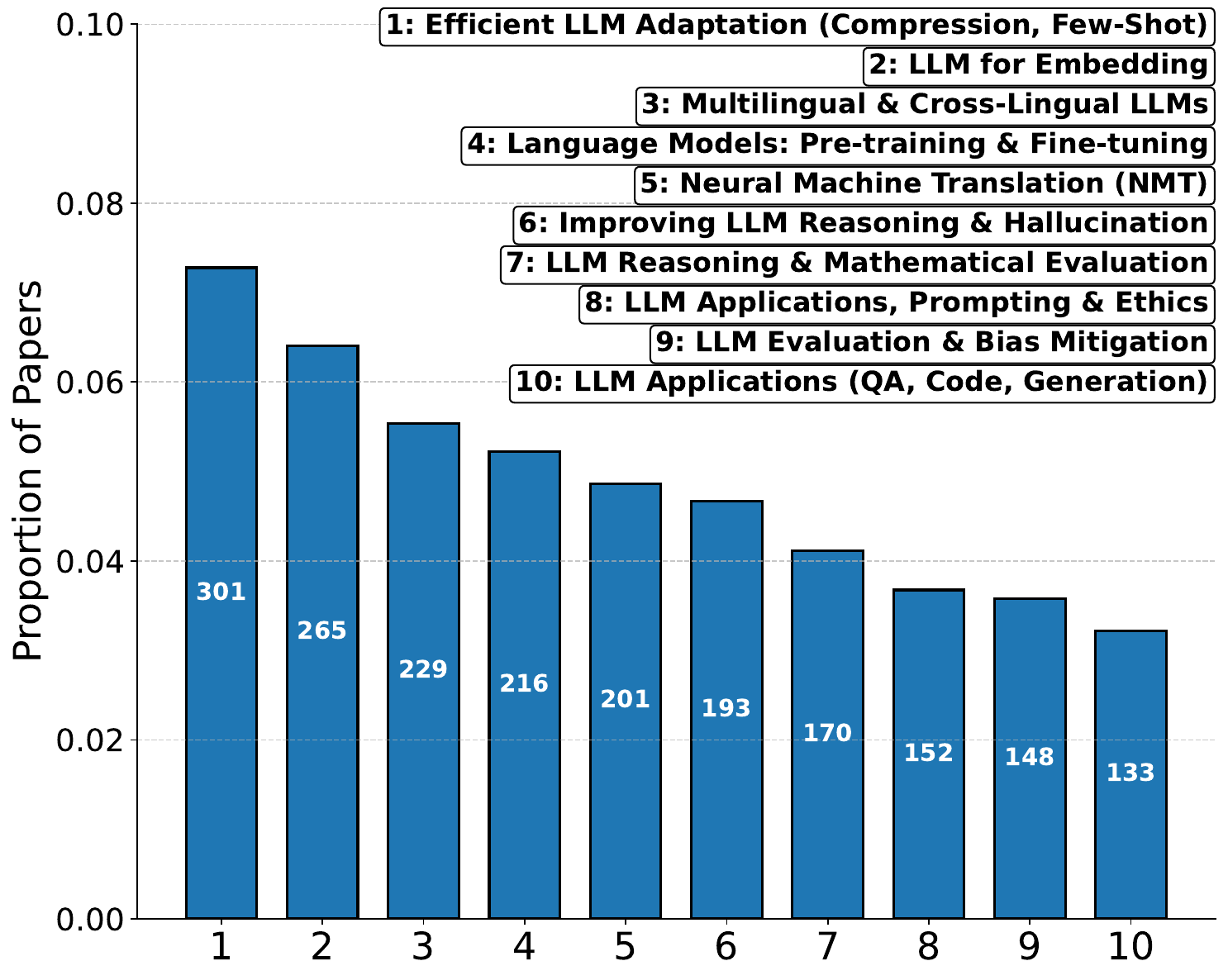}
        \vspace{-0.3in}
        \caption{Top 10 topics in EMNLP.}
        \label{fig:EMNLP}
    \end{minipage}
    \hspace{0.03in}
    \begin{minipage}[t]{.32\textwidth}
        \centering
        \includegraphics[width=\textwidth]{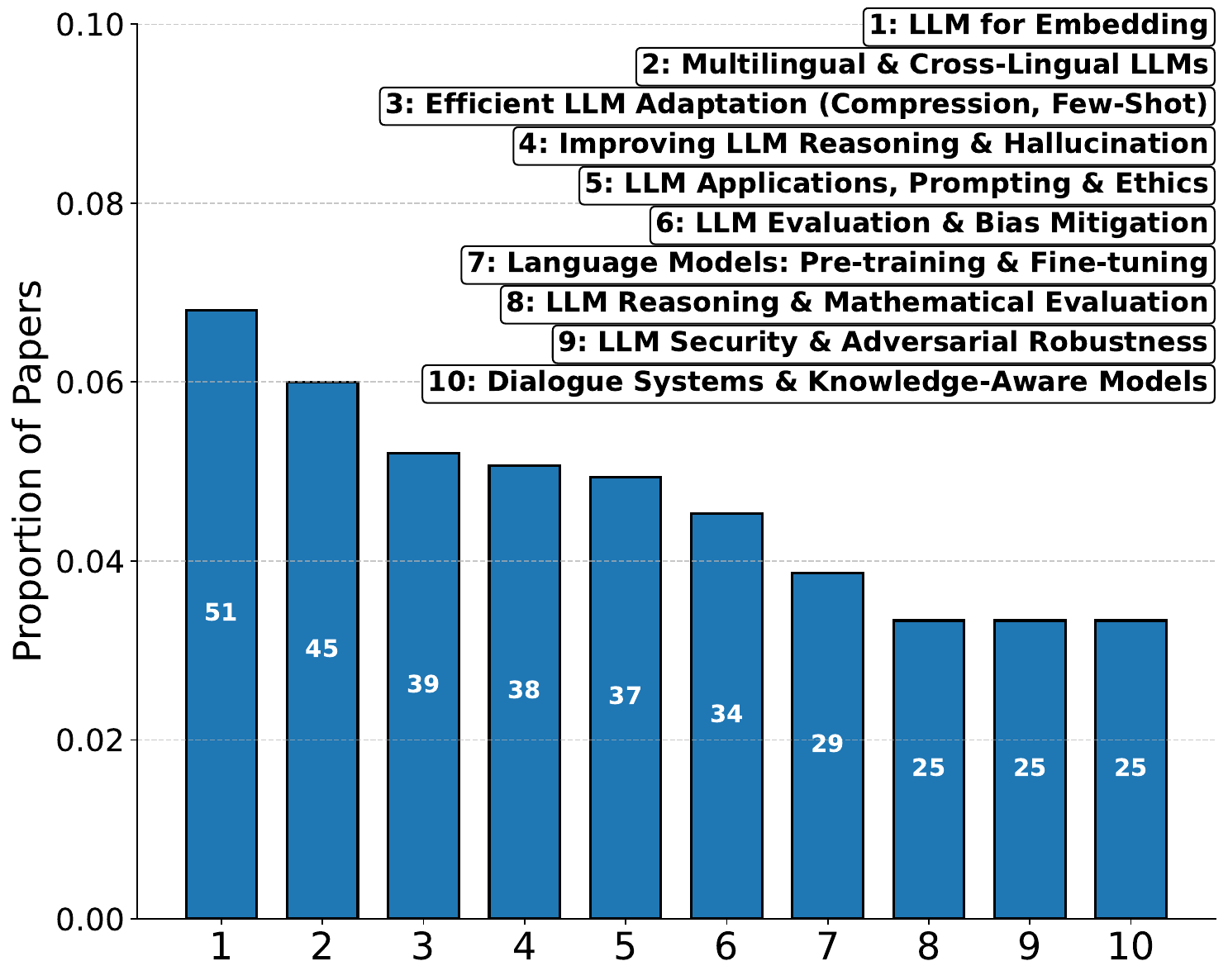}
        \vspace{-0.3in}
        \caption{Top 10 topics in NAACL.}
        \label{fig:NAACL}
    \end{minipage}

    \vspace{0.03in} 

    \begin{minipage}[t]{.32\textwidth}
        \centering
        \includegraphics[width=\textwidth]{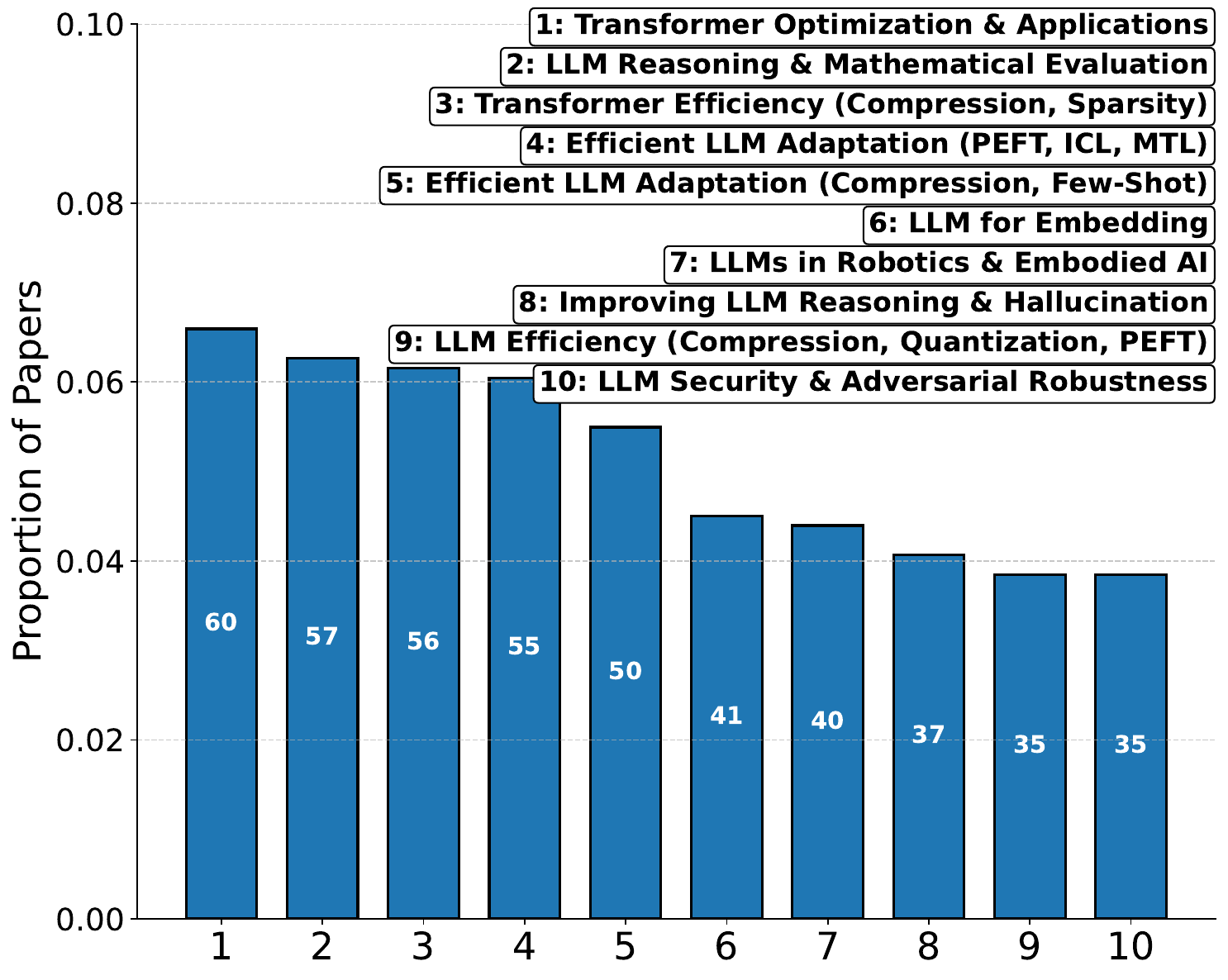}
        \vspace{-0.3in}
        \caption{Top 10 topics in ICLR.}
        \label{fig:ICLR}
    \end{minipage}
    \hspace{0.03in}
    \begin{minipage}[t]{.32\textwidth}
        \centering
        \includegraphics[width=\textwidth]{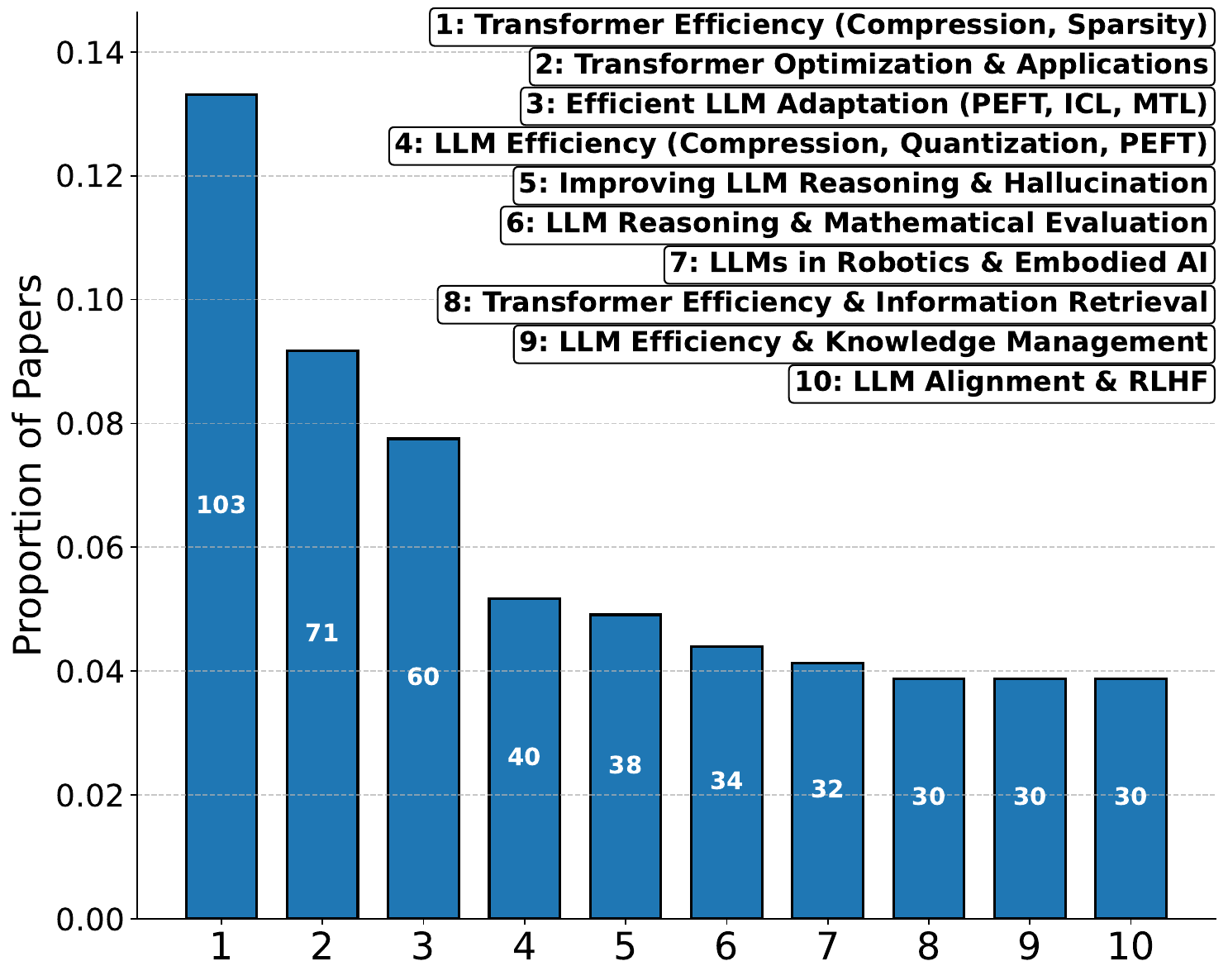}
        \vspace{-0.3in}
        \caption{Top 10 topics in ICML.}
        \label{fig:ICML}
    \end{minipage}
    \hspace{0.03in}
    \begin{minipage}[t]{.32\textwidth}
        \centering
        \includegraphics[width=\textwidth]{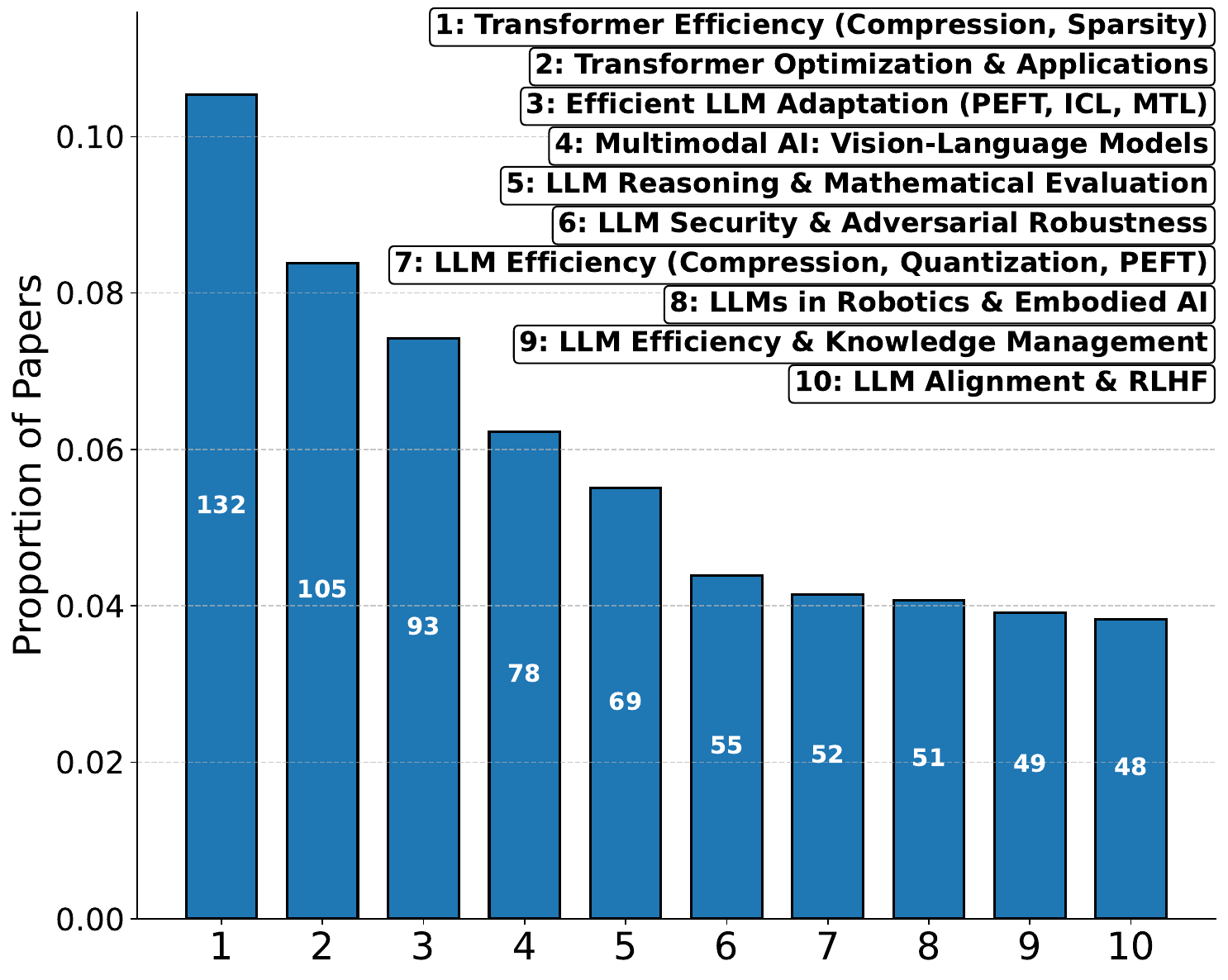}
        \vspace{-0.3in}
        \caption{Top 10 topics in NeurIPS.}
        \label{fig:NeurIPS}
    \end{minipage}

    \vspace{0.03in} 

    \begin{minipage}[t]{0.32\textwidth}
        \centering
        \includegraphics[width=\textwidth]{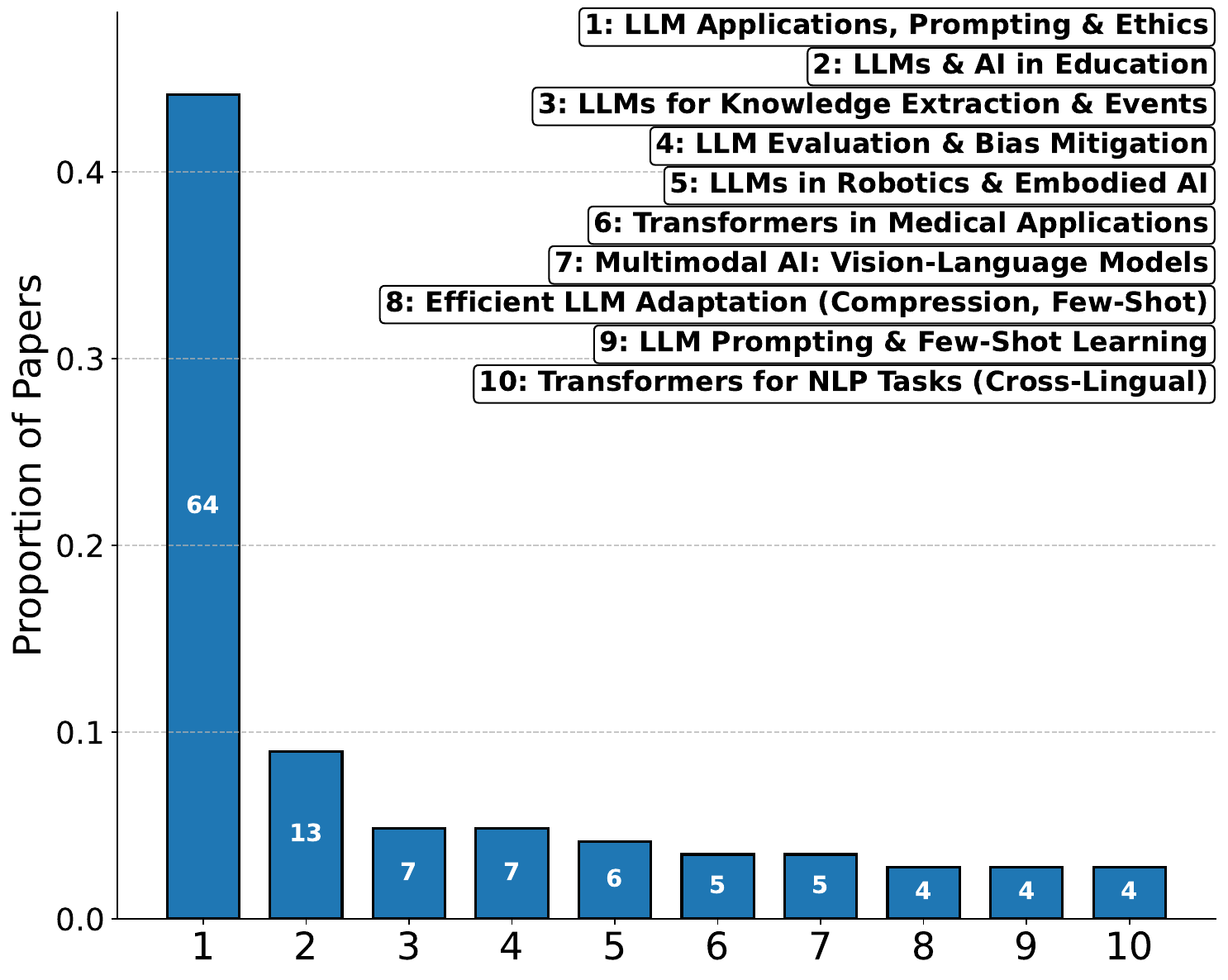}
        \vspace{-0.3in}
        \caption{Top 10 topics in CHI.}
        \label{fig:CHI}
    \end{minipage}
    \hspace{0.03in}
    \begin{minipage}[t]{0.32\textwidth}
        \centering
        \includegraphics[width=\textwidth]{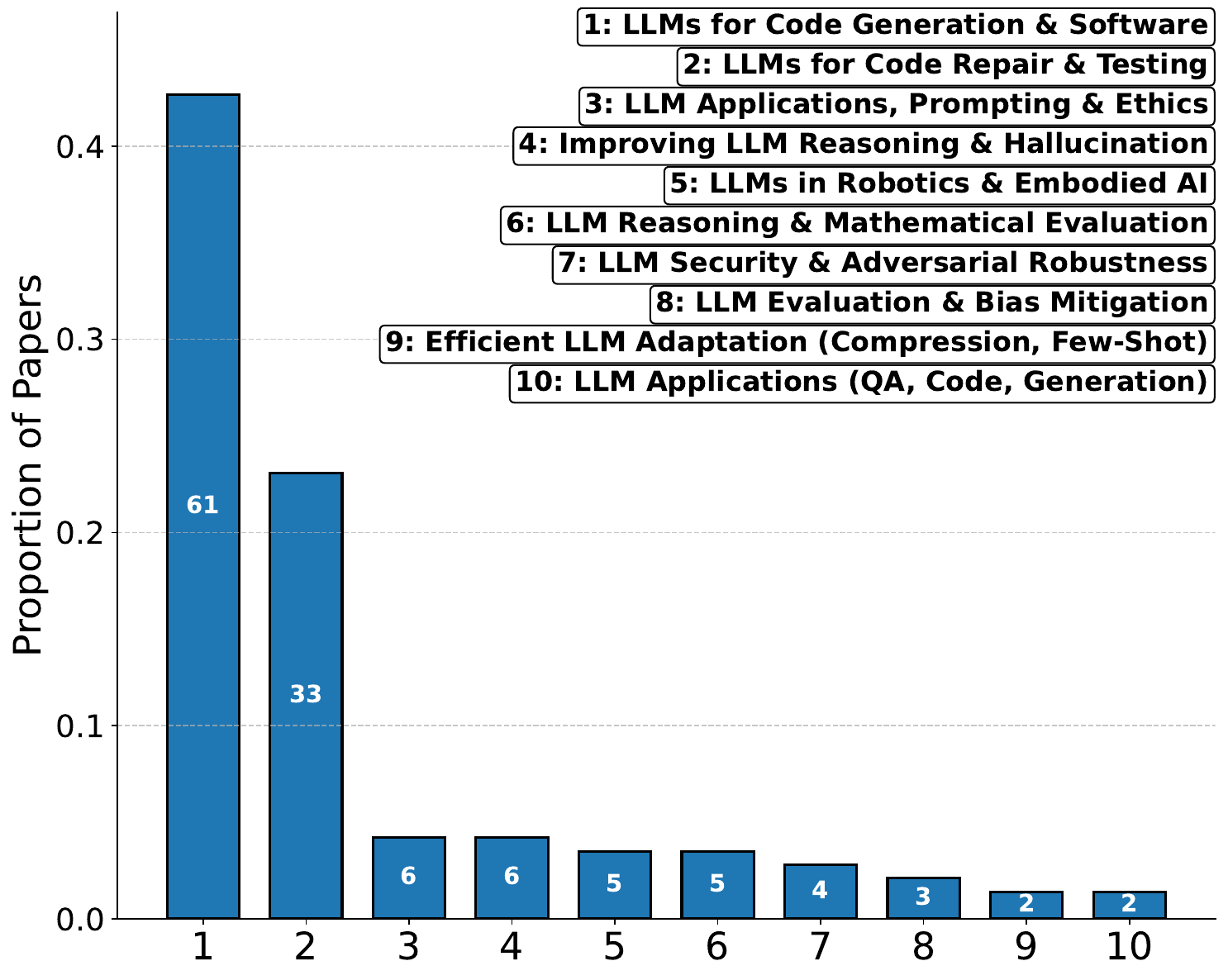}
        \vspace{-0.3in}
        \caption{Top 10 topics in ASE.}
        \label{fig:ASE}
    \end{minipage}
    \hspace{0.03in}
    \begin{minipage}[t]{0.32\textwidth}
        \centering
        \includegraphics[width=\textwidth]{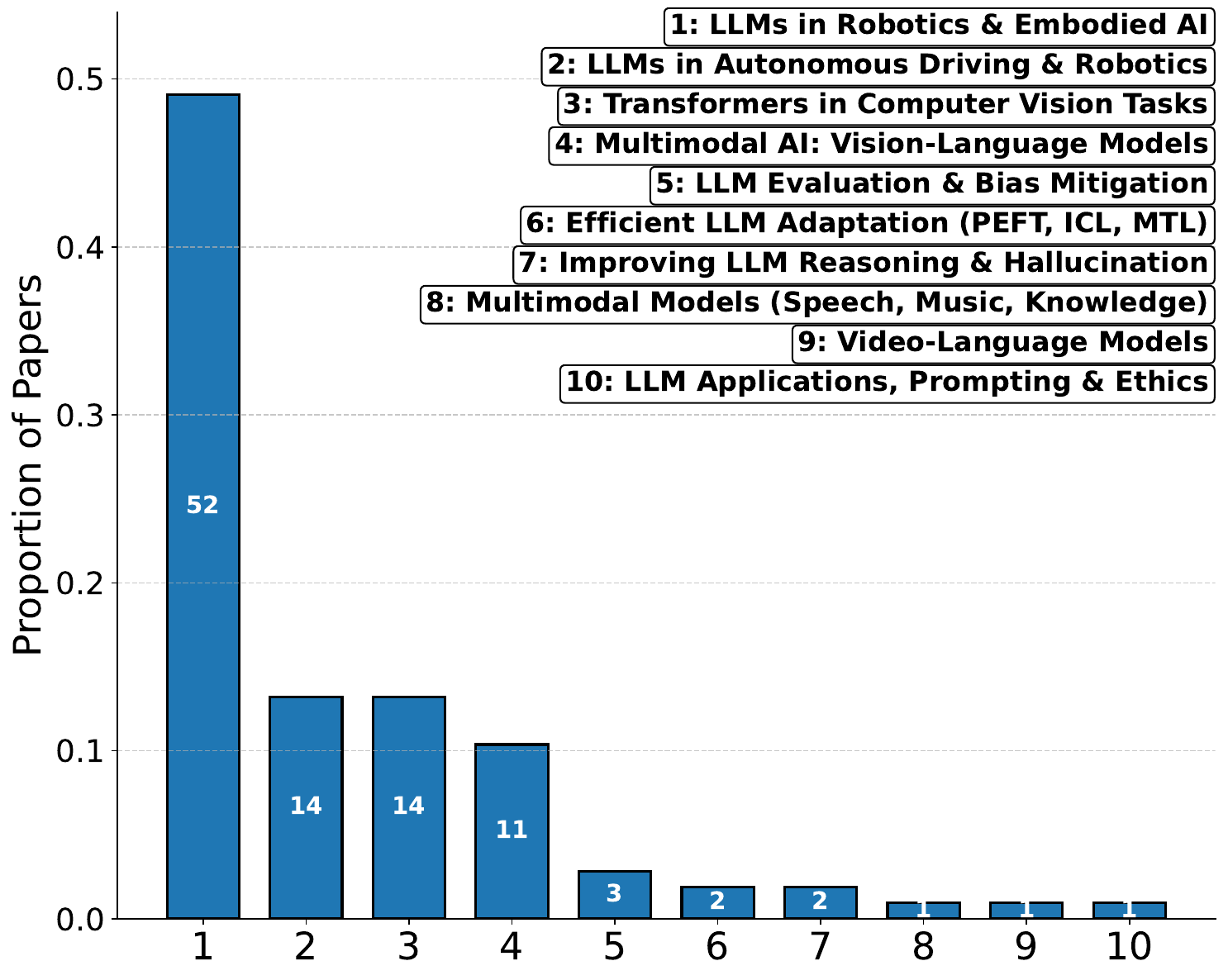}
        \vspace{-0.3in}
        \caption{Top 10 topics in ICRA.}
        \label{fig:ICRA}
    \end{minipage}
\end{figure*}

\textbf{Insight \#3. The interdisciplinary expansion trend underscores the broadening impact of LLMs on diverse domains like HCI, ubiquitous computing, and robotics.}
In Interdisciplinary Areas, the influence of LLMs is steadily growing. Conferences such as CHI, UIST, and UBICOMP enjoy a noticeable expansion in LLM-related papers, with proportions surpassing 10\% by 2024, reflecting the growing integration of LLMs in human-computer interaction (HCI) and related applications. Similarly, robotics conferences such as ICRA, IROS, and RSS are increasingly embracing more LLM-related papers, highlighting the tendency to equip robots with LLM to enhance robotic capabilities.

\subsection{Topic Analysis}
To understand the focal points and distribution of research topics surrounding LLMs, we analyze the top 10 topics of LLM-related papers across nine prominent conferences. This selection spans core NLP conferences (ACL, EMNLP, NAACL), major machine learning venues (ICLR, ICML, NeurIPS), and leading conferences in specialized domains like HCI (CHI), software engineering (ASE), and robotics (ICRA). Examining these diverse conferences provides a broad perspective on how LLM research is pursued, highlighting both common themes and domain-specific concentrations, as indicated by the proportion of papers dedicated to various topics in Fig.~\ref{fig:ACL} to Fig.~\ref{fig:ICRA}.

\textbf{Insight \#4. Leading NLP conferences focus heavily on LLM adaptation, evaluation, and core development, underscoring their central role in advancing LLM.}
Research presented at major NLP conferences (ACL, EMNLP, NAACL, Fig.~\ref{fig:ACL}-~\ref{fig:NAACL}) consistently spotlight research centered around the use of LLMs for embedding, evaluation, and efficient adaptation. Core LLM development practices such as pre-training, fine-tuning, and enhancement of reasoning abilities are also commonly explored. These conferences serve as the research hubs for advancing LLM research.

\textbf{Insight \#5. Machine learning conferences prioritize the architectural and efficiency improvements of LLMs, reflecting a foundational interest in scaling and optimizing Transformer-based models.}
At Machine Learning conferences (ICLR, ICML, NeurIPS, Fig.~\ref{fig:ICLR}-~\ref{fig:NeurIPS}), researchers pay more attention to optimizing the Transformer architecture, which is fundamental to many LLMs. A major emphasis is placed on enhancing Transformer and LLM efficiency through techniques like compression, sparsity, quantization, and parameter-efficient fine-tuning (PEFT). These efforts are often coupled with research on LLM adaptation and their integration into other tasks, such as robotics and CV. 

\textbf{Insight \#6. Application-driven conferences reveal the domain-specific adaptation of LLMs, emphasizing the integration of LLMs into real-world systems.}
At CHI (Fig.~\ref{fig:CHI}), the community emphasizes the practical application of LLMs, with a strong focus on prompting strategies and ethical considerations. In ASE (Fig.~\ref{fig:ASE}), the focus shifts to leveraging LLMs for automated software engineering tasks, including code generation, repair, and testing, as well as the broader application of LLMs in software development workflows. Meanwhile, ICRA (Fig.~\ref{fig:ICRA}) explores the integration of LLMs into robotics and embodied AI systems, with research addressing applications in multimodal models and autonomous driving, reflecting the field’s interest in bridging LLMs with robotic intelligence.

\begin{figure}[t]
\centering
\includegraphics[width=\columnwidth]{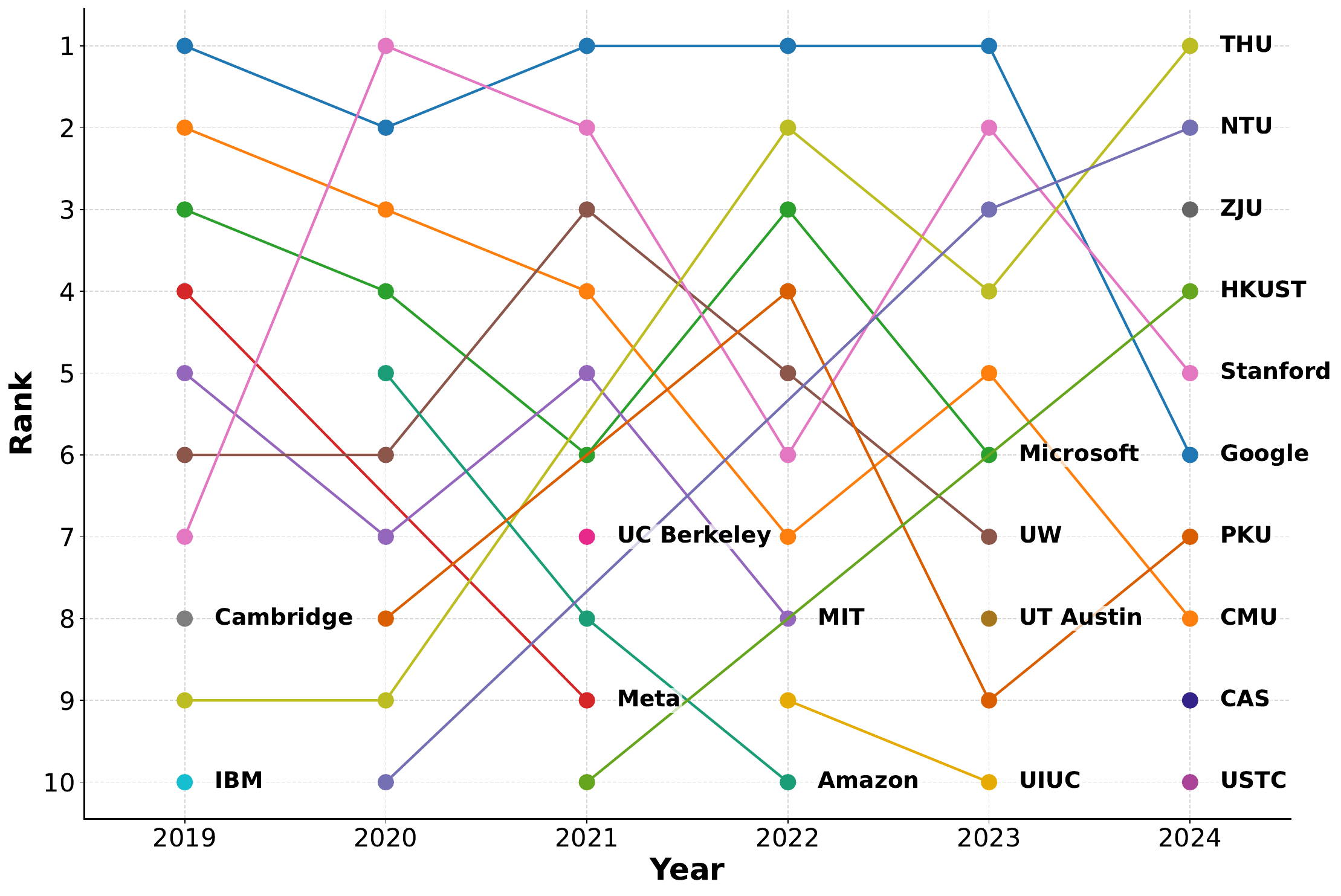}
\caption{Top 10 affiliations by LLM-related paper count.}
\label{fig:affiliation_trend}
\end{figure}

\subsection{Affiliation Analysis}
To figure out which industry or university is leading the research on LLM, we plot Fig.~\ref{fig:affiliation_trend}. It illustrates the shifting landscape of leadership in LLM research publications from 2019 to 2024, tracking the ranks of the top 10 performing affiliations based on publication volume.

\textbf{Insight \#7. Industry affiliations dominated the early phase of LLM research.}
From 2019 through the early 2020s, major technology companies such as Google, Microsoft, Meta, and Amazon led the field in terms of publication volume. Google, in particular, maintained either the No. 1 or 2 position from 2019 to 2023, reflecting its sustained contributions to foundational model development and deployment. This era of dominance was fueled by the industry’s access to vast computational resources, proprietary data, and a strong investment on the development of LLMs. 

\textbf{Insight \#8. Academic institutions have steadily gained ground, reflecting increased research capacity and influence.}
Over time, universities such as Tsinghua University (THU), Nanyang Technological University (NTU), Stanford, the University of Washington (UW), and the Hong Kong University of Science and Technology (HKUST) rose in rank, gradually challenging and overtaking their industry counterparts. This indicates a growing ability within academia to contribute high-impact research in LLMs, likely supported by rising collaboration with industry, access to open-source models, and a broader institutional focus on AI.
By 2024, THU, NTU, Zhejiang University (ZJU), and HKUST established themselves as leading entities in the LLM research space, which highlights a broader trend of increasing academic leadership and, as industry contributions, though still significant, become comparatively less dominant in shaping the trajectory of LLM research.

\subsection{Nation Analysis}

\begin{figure}[t]
\centering
\includegraphics[width=\columnwidth]{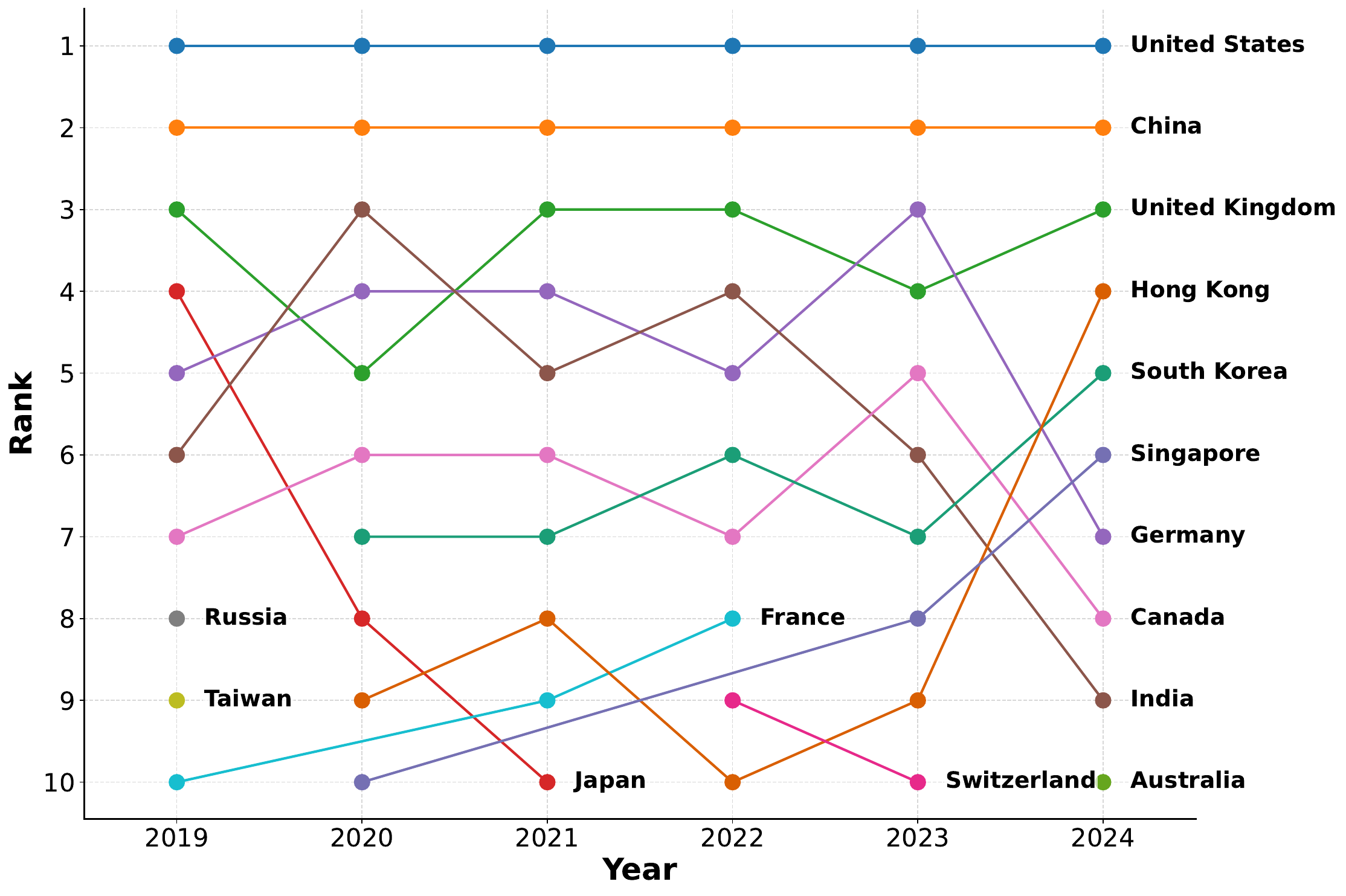}
\caption{Top 10 nations/dependencies by LLM-related paper count.}
\label{fig:nation_trend}
\end{figure}

To understand the global landscape and evolving geographic distribution of LLM research, we examine national-level contributions over time. Fig.~\ref{fig:nation_trend} illustrates the ranking trends of the top 10 nations/dependencies based on their publication counts of LLM-related papers from 2019 to 2024.

\textbf{Insight \#9. The global landscape of LLM research is marked by stable leadership of the United States and China, alongside the rising prominence of emerging contributors.}
The rankings highlight remarkable stability at the top, with the United States and China consistently occupying the first and second positions, respectively, reflecting sustained global dominance in LLM research output. The United Kingdom also repeatedly ranks third, briefly dropping to fourth in 2023 but returning to third by 2024. A significant development is Hong Kong's notable advancement, progressively rising to reach the fourth position by 2024, highlighting its increasing influence in LLM. South Korea, Singapore, Germany, India, and Canada continuously maintain stable positions within the top ten rankings, underscoring their steady and significant contributions to LLM research throughout the observed period.

\begin{figure*}[tbp]
    \centering
    \begin{minipage}[t]{.32\textwidth}
        \centering
        \includegraphics[width=\textwidth]{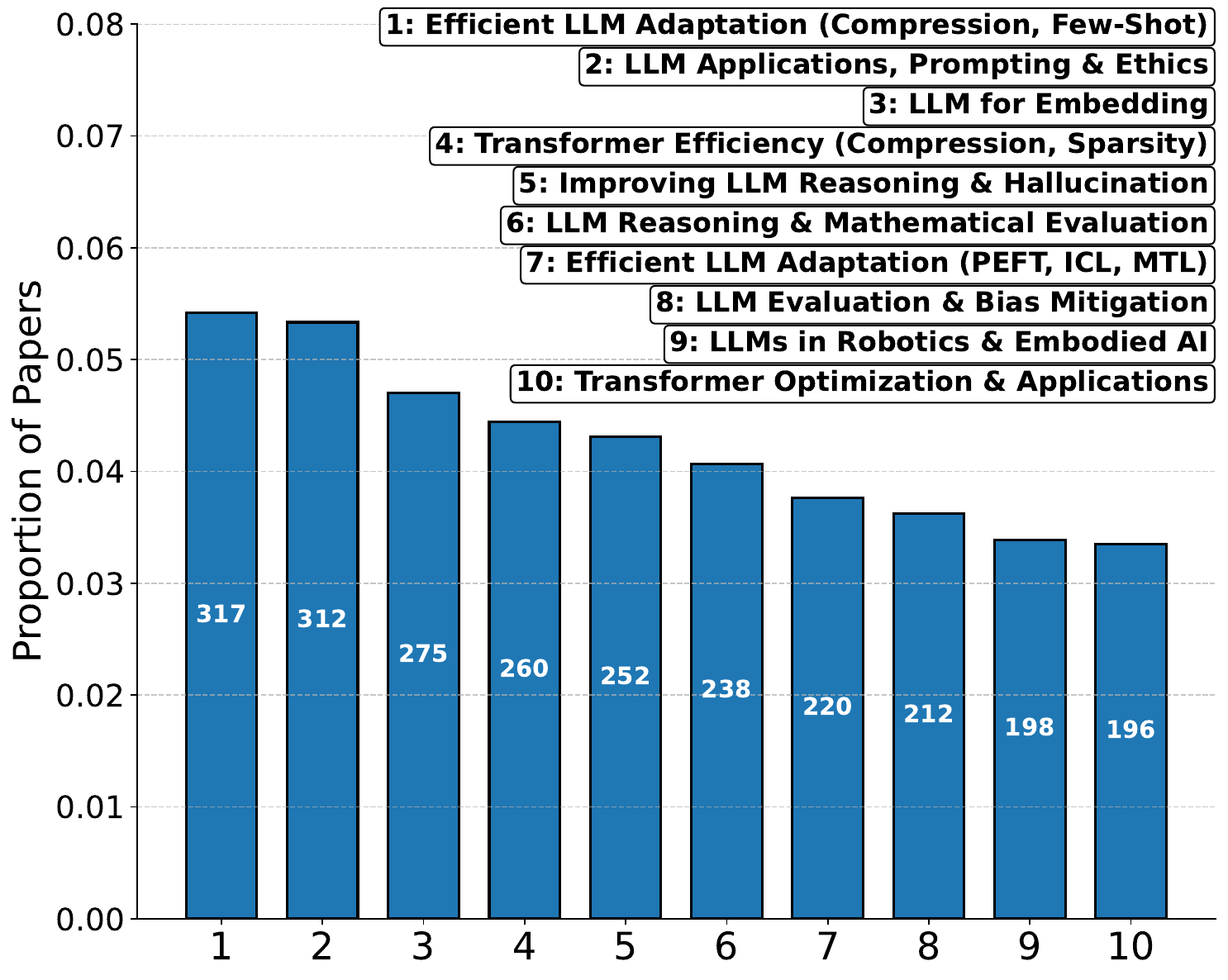}
        \vspace{-0.3in}
        \caption{Top 10 topics in the United States.}
        \label{fig:USA}
    \end{minipage}
    \hspace{0.03in}
    \begin{minipage}[t]{.32\textwidth}
        \centering
        \includegraphics[width=\textwidth]{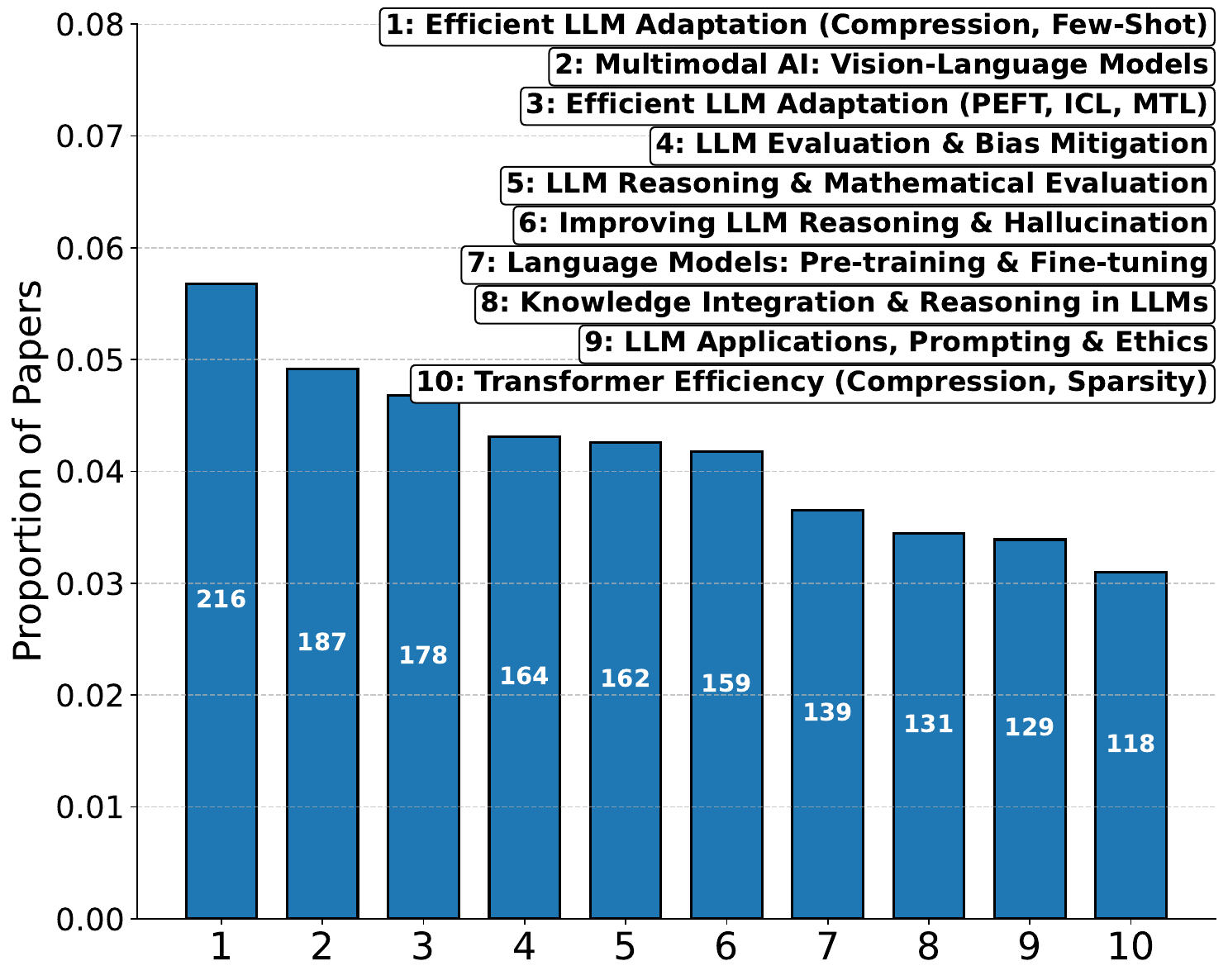}
        \vspace{-0.3in}
        \caption{Top 10 topics in China.}
        \label{fig:China}
    \end{minipage}
    \hspace{0.03in}
    \begin{minipage}[t]{.32\textwidth}
        \centering
        \includegraphics[width=\textwidth]{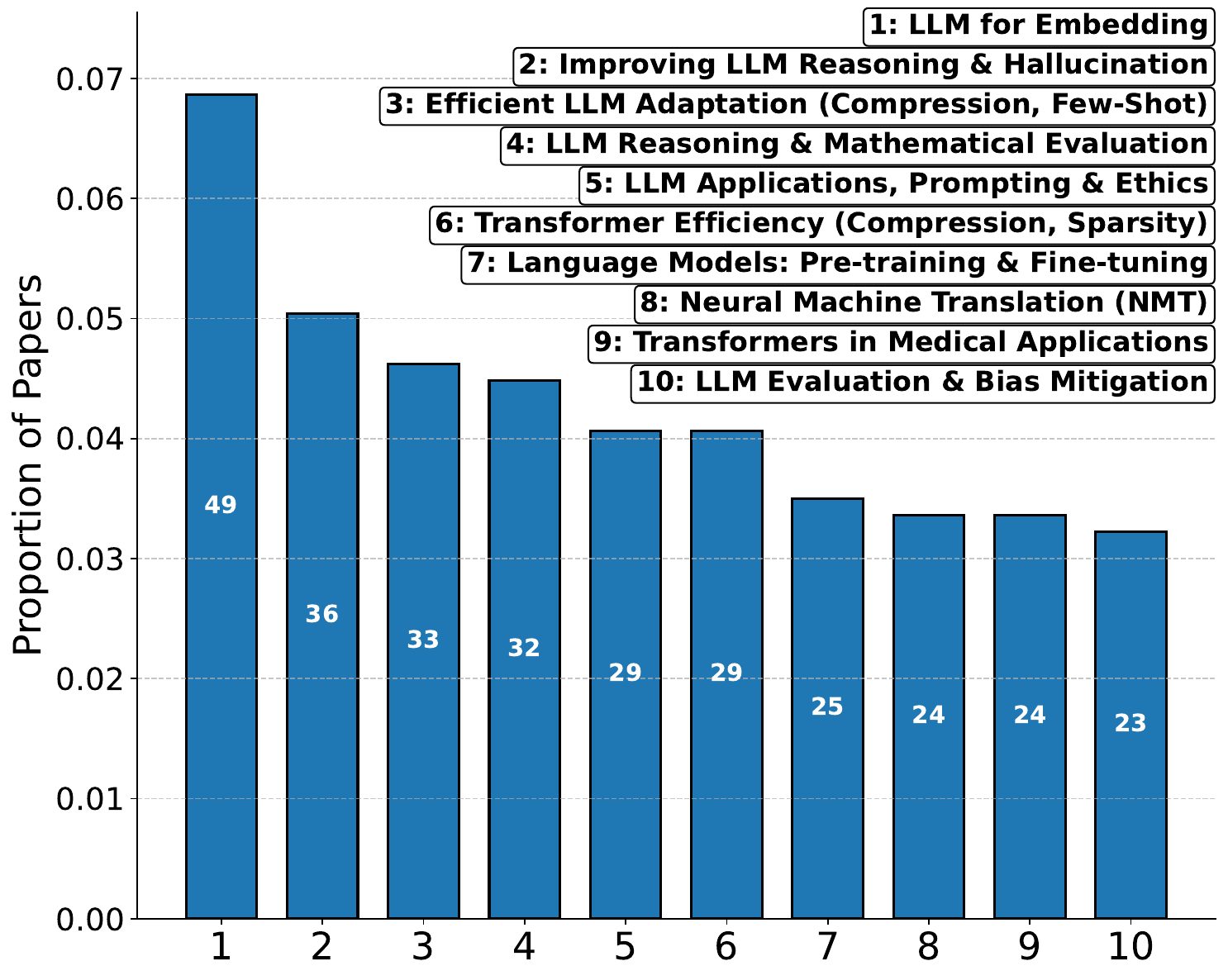}
        \vspace{-0.3in}
        \caption{Top 10 topics in the United Kingdom.}
        \label{fig:UK}
    \end{minipage}
\end{figure*}

To further explore these national research dynamics, we now analyze the distribution of topics within the top three nations— the United States, China, and the United Kingdom. 

\textbf{Insight \#10. Leading nations share core LLM research priorities while exhibiting distinct thematic specializations.}
The analysis of the top 10 LLM research topics across the United States, China, and the United Kingdom (Fig.~\ref{fig:USA}-\ref{fig:UK}) reveals both significant overlaps and distinct national priorities. Core areas such as efficient LLM adaptation, LLM reasoning, and improving LLM hallucination garner substantial attention in all three nations, indicating shared foundational research interests and goals. However, thematic specializations are evident. For instance, while the US places a strong emphasis on LLM application, prompting, and LLM for embedding, China demonstrates a particular focus on vision-language models. Conversely, the UK prioritizes ``LLM for Embedding'' most highly. We hypothesize that unique national interests emerge in specific application domains, such as robotics in the US, multimodal systems in China, and neural machine translation and medical applications in the UK, could be the root cause of such diverse focuses. 

\vspace{-.1in}
\section{Conclusion}~\label{sec:conclusion}
This paper presents ten important insights related to LLM research. Particularly, 
this manuscript comprehensively analyzed over 16,193 LLM-related papers from 77 top-tier computer science conferences, highlighting the evolving interdisciplinary landscape of LLM research. The study identifies significant shifts in research focus influenced by LLM advancements, underscores the rising prominence of academic institutions alongside industry, and reveals distinct national patterns shaping global research trends. These insights provide a foundation for guiding future research directions and policy-making in the rapidly evolving field of LLMs. 

\vspace{-.08in}
\bibliographystyle{IEEEtran}
\bibliography{Reference}

\end{document}